\documentclass[aps,amsmath,amssymb,prb,twocolumn,groupedaddress,superscriptaddress,showpacs]{revtex4-2}

\usepackage{graphicx}

\begin{document}

\title{Strong negative magnetoresistance and hopping transport in graphenized nematic aerogels}

\author{V.I. Tsebro}
\email[]{v.tsebro@mail.ru}
\affiliation{P.N. Lebedev Physical Institute, RAS, 53 Leninsky Prospect, 119991 Moscow, Russia}
\affiliation{P.L. Kapitza Institute for Physical Problems, RAS, 2 Kosygina Street, 119334 Moscow, Russia}

\author{E.G. Nikolaev}
\affiliation{P.L. Kapitza Institute for Physical Problems, RAS, 2 Kosygina Street, 119334 Moscow, Russia}

\author{M.S. Kutuzov}
\affiliation{Metallurg Engineering Ltd., Tallinn, 11415 Estonia}

\author{A.V. Sadakov}
\affiliation{P.N. Lebedev Physical Institute, RAS, 53 Leninsky Prospect, 119991 Moscow, Russia}

\author{O.A. Sobolevskiy}
\affiliation{P.N. Lebedev Physical Institute, RAS, 53 Leninsky Prospect, 119991 Moscow, Russia}


\begin{abstract}
The transport properties of nematic aerogels, which consist of oriented mullite nanofibers coated with a graphene shell, were studied. It is shown that the magnetoresistance of this system is well approximated by two contributions - negative one, described by the formula for systems with weak localization , and positive contribution, linear in the field and unsaturated in large magnetic fields. The behavior of phase coherence length on temperature obtained from the analysis of the negative contribution indicates the main role of the electron-electron interaction in the destruction of phase coherence and, presumably, the transition at low temperatures from a two-dimensional weak localization regime to a one-dimensional one. The positive linear contribution to magnetoresistance is apparently due to the inhomogeneous distribution of the local carrier density in the conductive medium. It has also been established that the temperature dependence of the resistance for graphenized aerogels with a low carbon content, when the graphene coating is apparently incomplete, can be represented as the sum of two contributions, one of which is characteristic of weak localization, and the second is described by hopping mechanism corresponding to the Shklovskii-Efros law in the case of a granular conductive medium. For samples with a high carbon content, there is no second contribution.
\end{abstract}


\maketitle

\section{Introduction}

Graphenization by chemical vapor deposition (CVD) of nematic aerogel based on $\gamma$-Al$_2$O$_3$ (naphen) or on aluminum silicate Al$_2$O$_3\cdot$SiO$_2$ (mullite), in the process during which its nanofibers are covered with one or more layers of graphene with a large number of defects \cite{hussainova2015,ivanov2016,solod2019}, makes aerogel conductive with very interesting properties. In Ref.~\cite{tsebro2022}, we characterized
graphenized mullite samples, in particular, we investigated their structure by scanning electron microscopy, determined the carbon content by X-ray photoelectron spectroscopy (XPS) to estimate the thickness of the graphene coating, and obtained Raman spectra, which allowed us to qualitatively assess the degree of defectivity and grain size of the carbon coating. As a result of measurements of the transport properties of bulk aerogel samples, the resistivity value and its anisotropy were determined. It was also shown that the temperature dependences of the electrical resistance of both bulk and compact samples of graphenized mullite in the range of 9--40~K can be described by Eq.~(\ref{eq:mott}) for variable range hopping (VRH) conductivity, in which the exponent $\alpha$ varies from 0.4 to 0.9 when the number of layers in the graphene nanofiber shell decreases from 4--6 to 1--2.

\begin{equation}\label{eq:mott}
    R(T)=R(0)\exp \left[ \left(\frac{T_0}{T}\right)^\alpha\, \right].
\end{equation}

The magnetoresistance measured in Ref.~\cite{tsebro2022} at 4.2~K in the magnetic fields up to 2.3~T was negative, significant in magnitude (MR =$(R(B)-R(0))/R(0)\approx -0.1$ in the field 2~T) and increasing as the thickness of the graphene shell decreases. Approximation of the field dependence of the magnetoresistance by the corresponding expression for weak localization of carriers in the two-dimensional case (2D-WL) allowed us to determine the values of the phase coherence length (13-15~nm), which are in a reasonable ratio with the size of graphene grains in the aerogel nanofiber shell. Thus, it turned out that this system exhibits properties characteristic of both strong localization (hopping conductivity) and weak localization (negative magnetoresistance). This circumstance, as well as the limited temperature range in which the law (\ref{eq:mott}) was satisfied, and especially the obtained low values of $T_0$ (below 30K) indicated the need for a more detailed study of the transport properties of graphenized nematic aerogel in a wide range of temperatures and magnetic fields.

In the present work, the resistance dependences of graphenized mullite in magnetic fields up to 16 T and in the temperature range from 3 to 300 K have been investigated. It was found that the magnetoresistance can be represented as the sum of two contributions: negative (MR$^-<0$) and positive (MR$^+>0$). For samples with a small number of layers in the graphene nanofiber shell, the negative contribution at temperatures near helium can reach $\approx -0.$5 in magnetic field of 16~T. It also turned out that the negative contribution of MR$^-$ in the whole region of magnetic fields and temperatures is described with good accuracy by the expression for the case of 2D weak localization \cite{hikami80,lee85}. The positive contribution, which is linear and unsaturated, seems to be due to the significant inhomogeneity of the conductive medium \cite{rama2017,ping2014,naray2015,zhu2022,friedman2010,kisslinger2015,gu2021}. The analysis of the temperature dependence of zero-field resistance has shown that in aerogel samples with a minimum number of layers (1-3) in the graphene nanofiber shell $R(T)$ contains two contributions: (a) the contribution $R_{\textrm{WL}}\propto  1/\ln T$ associated with weak localization under conditions of diffusive transport of carriers, and (b) the contribution $R_{\textrm{hop}}$ due to hopping mechanism according to Eq.~(\ref{eq:mott}). In this case, regardless of the graphene coating thickness, the value of $\alpha$ in Eq.~(\ref{eq:mott}) is equal to 1/2 (Efros-Shklovsky law), and $T_0$ = 210--260~K. For samples with a larger graphene shell thickness, the hopping contribution is absent.

\section{Aerogel samples and experimental details}

As previously \cite{tsebro2022}, samples with carbon content
14~at.\% (AG-14), 20~at.\% (AG-20), 31~at.\% (AG-31) and 44~at.\% (AG-44)
were studied. As it was noted in this work, if, based on the data on the carbon content in the samples and the average diameter of nanofibers, to estimate the thickness of the graphene shell, it turns out that it is 1-2 graphene layers for samples with the minimum carbon content (AG-14) and 4-6 layers for samples with the maximum content (AG-44).

The magnetoresistance measurements were carried out using the standard four-terminal method on individual fragments of aerogel with a large ratio of length to its transverse size. The cross section of such fragments was of the order of 0.05 mm$^2$ with a length of 6-8 mm.
\begin{figure}[h]
\begin{center}
\includegraphics[width=8cm]{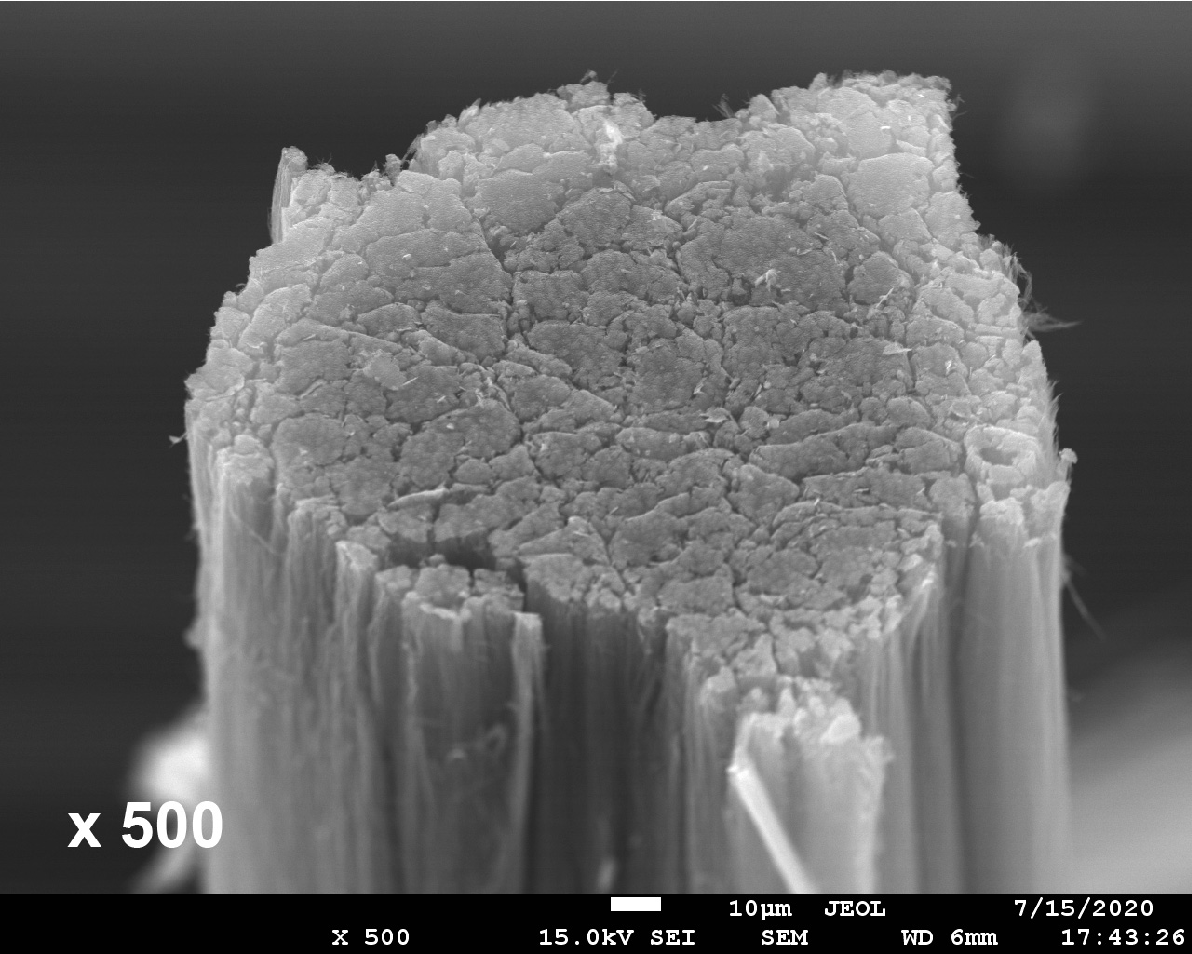}
  \caption{Electron microscopic image of the end face of an AG-44 aerogel fragment with a transverse dimension of about 0.2~mm.}
  \label{fig:SEM_x500}
\end{center}
\end{figure}
Fig.~1 shows an electron microscopic image of the end face of one of such fragments with a transverse dimension of about 0.2~mm. For each sample, a separate special holder (Fig. 2) was made of tinned copper wires with a diameter of 0.06~mm stretched at a small angle to the plane of the holder base, which served as leads to current and potential contacts to the sample. After placing the sample in the space between the contact wires and the plane of the holder base, a small drop of conductive self-hardening silver paste was applied from the back side of the contact wires in the place of contact to form a stable contact. It is clear that in such geometry the measured value of electrical resistance is determined mainly by its longitudinal (along the nanofiber direction) component, and when the magnetic field vector is directed normal to the sample axis, the measured magnetoresistance will be transverse.
\begin{figure}[h]
\begin{center}
\includegraphics[width=7cm]{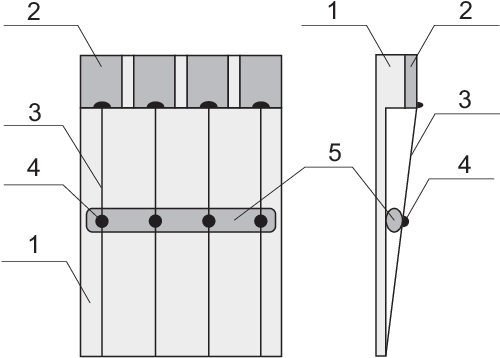}
  \caption{Design of aerogel sample holder in the form of nanofiber
fragment. 1 -- holder base (glass-textolite); 2 -- contact pads (copper); 3 -- contact wires (copper wire); 4 -- self-hardering conductive silver paste; 5 -- nanofiber aerogel fragment.}
  \label{fig:holder}
\end{center}
\end{figure}

\section{Experimental results and discussion}

\subsection{Magnetoresistance}

Fig.~3 shows the MR$(B)$ dependences measured at different temperatures for sample AG-14 with the lowest carbon content (number of graphene layers in the nanofiber shell 1--2), and in Fig.~4 MR$(B)$ dependences for samples AG-31 and AG-44 with twice or three times greater thickness of the graphene shell.

\begin{figure}[h]
\begin{center}
\includegraphics[width=8cm]{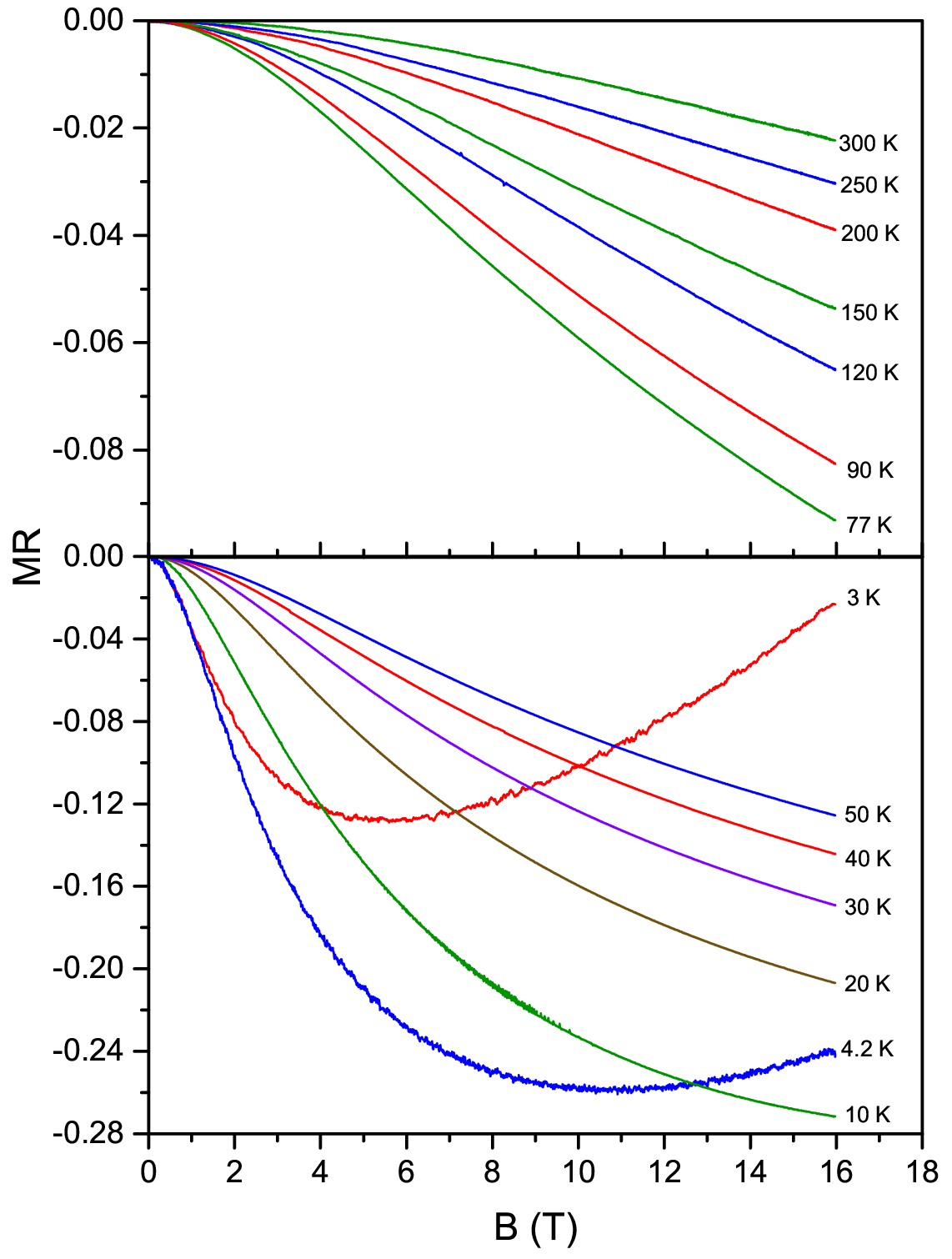}
  \caption{Magnetoresistance of AG-14 sample at temperatures from 3 to 300~K.}
  \label{fig:AG-14_all_T}
\end{center}
\end{figure}

\begin{figure}[h]
\begin{center}
\includegraphics[width=8cm]{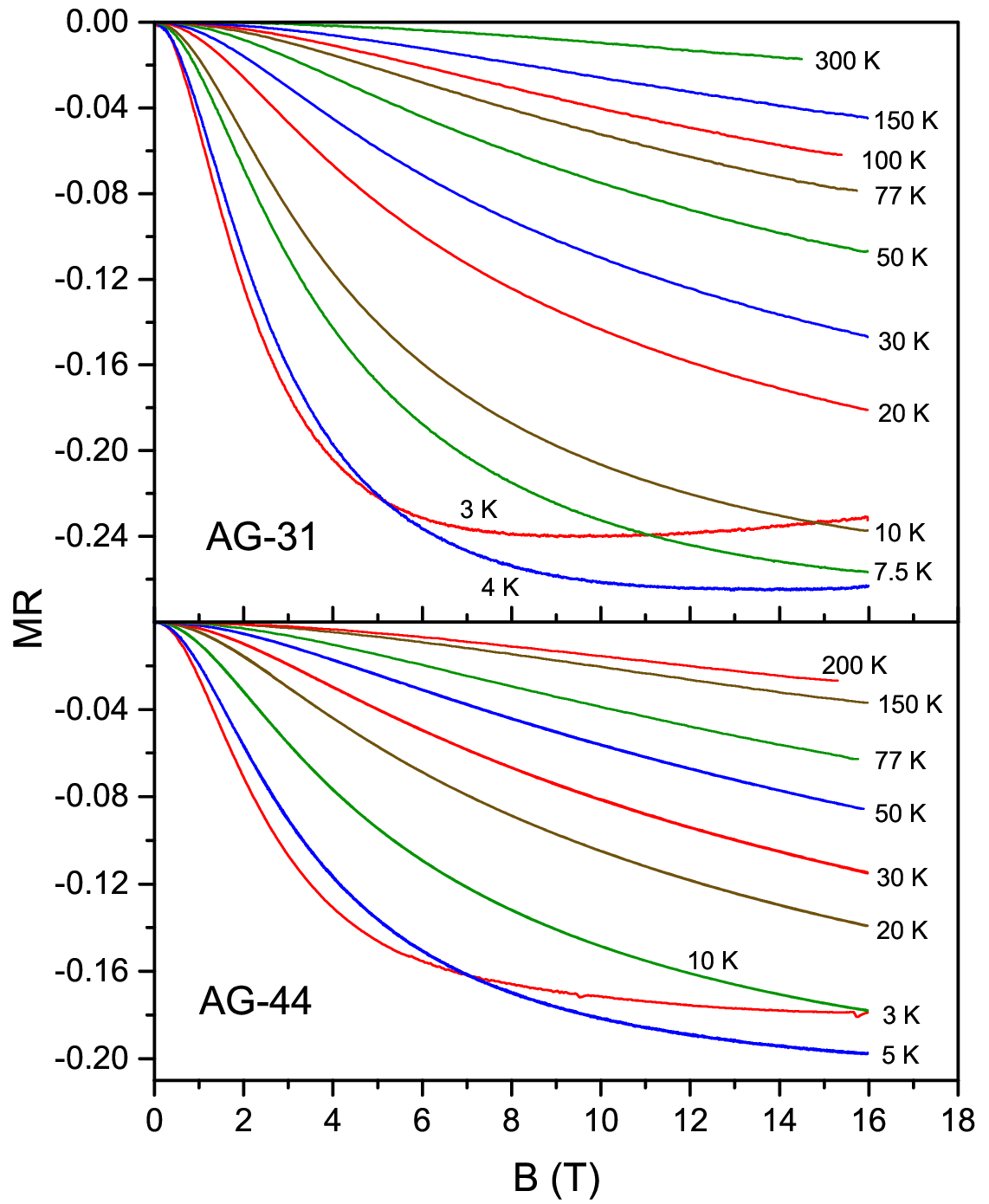}
  \caption{Magnetoresistance of samples AG-31 and AG-44 at different temperatures.}
  \label{fig:AG-31-44_all_T}
\end{center}
\end{figure}

It can be seen that at all temperatures the magnetoresistance is negative, significant in magnitude, and decreases by only an order of magnitude when the temperature changes from helium to room temperature. It can also be seen that at low temperatures a noticeable positive contribution to the magnetoresistance appears, leading to a strong deformation of the
MR$(B)$ dependences at temperatures 3--5 K.

When processing the experimental data, it turned out that at all temperatures the field dependences of MR are divided with good accuracy into two contributions: a negative ${\rm MR}^-<0$ and a smaller positive ${\rm MR}^+>0$, which, as it turned out, is strictly linear in the magnetic field in fields larger than 1~T, and quadratic in the field at $B\rightarrow 0$. At the same time, the negative contribution at all temperatures can be described using the well-known expression for the magnetocondensation MC in the case of 2D-WL \cite{hikami80,lee85,aleiner99} as follows:
\begin{equation}\label{eq:MR_MC}
    {\rm MR}^-(B,T)=\frac{1}{{\rm MC}(B,T)+1}-1\ ,
\end{equation}
where
\begin{multline}\label{eq:MC_WL}
{\rm MC}(B,T) = \frac{G(B,T)-G(0,T)}{G(0,T)}= {} \\
{} =A(T)\left[\Psi\left(\frac{1}{2}+\frac{B_{\phi}(T)}{B}\right)+
    \ln\left(\frac{B}{B_{\phi}(T)}\right)\right]{}\ .
\end{multline}
Here, the conductance $G(B,T) = 1/R(B,T)$, $A(T)$ is a temperature-dependent constant determined by the system parameters, $\Psi$ is the digamma function, $$B_{\phi}(T)=\hbar/4eL_{\phi}^2\ ,$$ $L_{\phi}$ --- phase coherence length.

The procedure of experimental data processing to separate the contributions of ${\rm MR}^-<0$ and ${\rm MR}^+>0$ was carried out in two stages and was summarized as follows. First, we approximated the dependences of magnetoresistance on magnetic field measured at a given temperature in the field range from 1 to 16 T by the sum of
\begin{equation}\label{eq:MR_summ}
    {\rm MR}(B,T)={\rm MR}^-(B,T)+{\rm MR}^+(B,T)\,,
\end{equation}
with three temperature-dependent parameters: $A, B_{\phi}$ of Eq.~(\ref{eq:MC_WL}), and the coefficient $C$ of positive linear contribution in magnetic fields above 1~T. That is, in this sum ${\rm MR}^-<0$ is expressed by formulas Eqs.~(\ref{eq:MR_MC},\ref{eq:MC_WL}), and ${\rm MR}^+>0$ by the formula
\begin{equation}\label{eq:MR_plus}
    {\rm MR}^+(B,T)=C(T)B\ .
\end{equation}

Moreover, since graphene grains in the aerogel nanofiber shell are located at different angles to the direction of the magnetic field, and the destruction of the contribution to the magnetoresistance associated with weak localization occurs due to the magnetic field component perpendicular to the graphene plane, the approximation was averaged over all possible angles according to the expression
\begin{equation*}
\left[\frac{\Delta R(B)}{R(0)}\right]
= \frac{1}{\pi} \int_{-\pi/2}^{\pi/2}\left[\frac{\Delta R(B\cos\theta)}{R(0)}\right]d\theta
\end{equation*}
as it was done, for example, in Ref.~\cite{piraux2015}  for highly conductive carbon nanotube fibers.

\begin{figure}[h]
\begin{center}
\includegraphics[width=8cm]{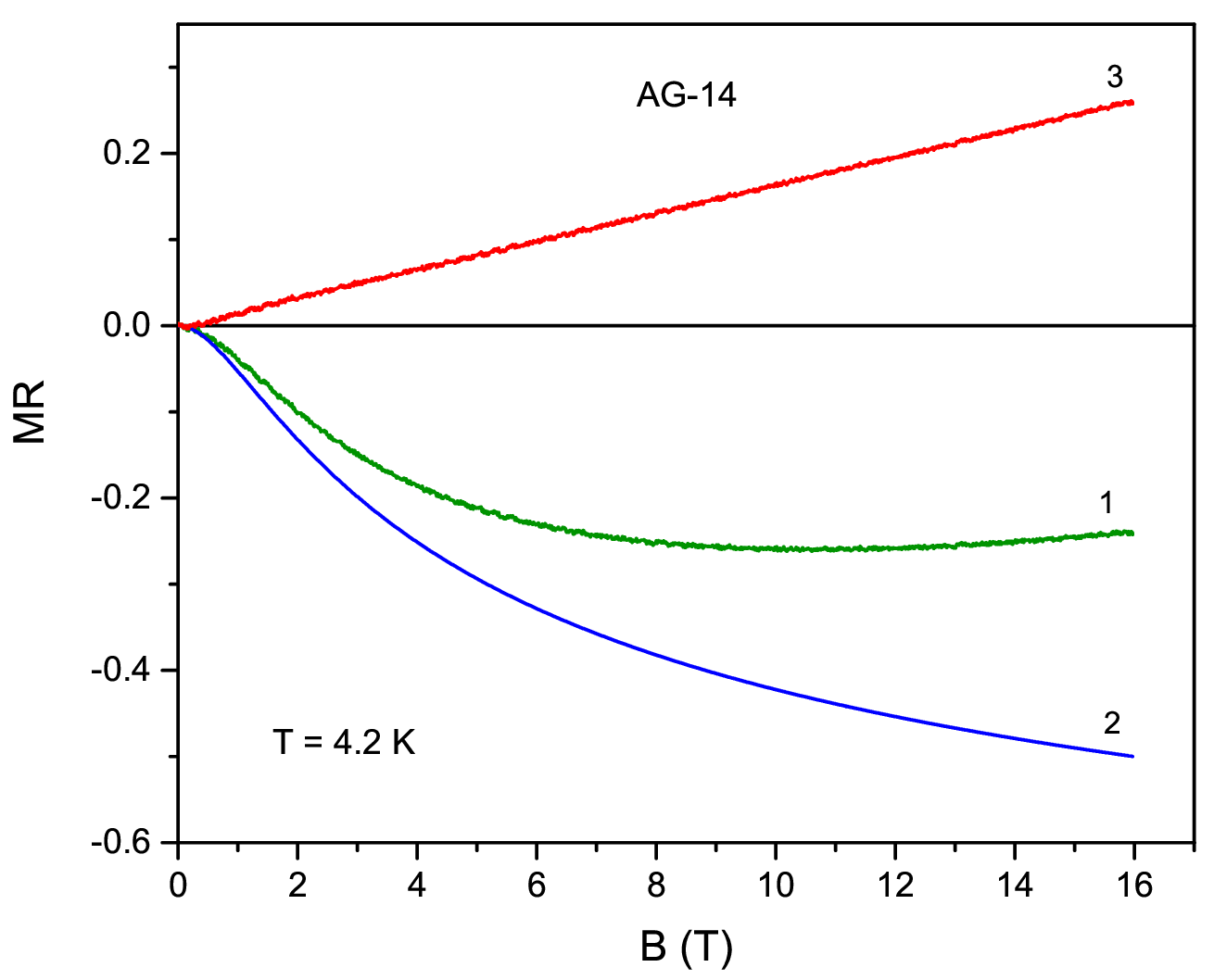}
  \caption{An example of the separation of contributions to the magnetoresistance for sample AG-14 at $T=4.2$~K. 1 -- experimental dependence of the magnetoresistance on the magnetic field, 2 -- dependence on $B$ negative contribution MR$^-$ according to Eqs.~(\ref{eq:MR_MC},\ref{eq:MC_WL}), and 3 -- dependence on $B$ positive contribution MR$^+$.}
  \label{fig:AG-14_T4K_3in1}
\end{center}
\end{figure}

At the second stage, from the difference of experimental MR values and MR$^-$ values, calculated by Eqs.~(\ref{eq:MR_MC},\ref{eq:MC_WL}) with parameters $A$ and $B_{\phi}$, obtained at the first stage of processing, the type of magnetic field dependence of the positive contribution ${\rm MR}^+$ was determined. Fig.~\ref{fig:AG-14_T4K_3in1} shows, as an example, the result of the separation of the contributions to the magnetoresistance for sample AG-14 at $T=4.2$~K. It can be seen that both contributions, negative and positive, differ in high fields in absolute value by a factor of about two, so that in a maximum magnetic field of 16 T, in the absence of the positive contribution, the magnetoresistance of this sample would reach a value of approximately -0.5.

\begin{figure}[h]
\begin{center}
\includegraphics[width=8cm]{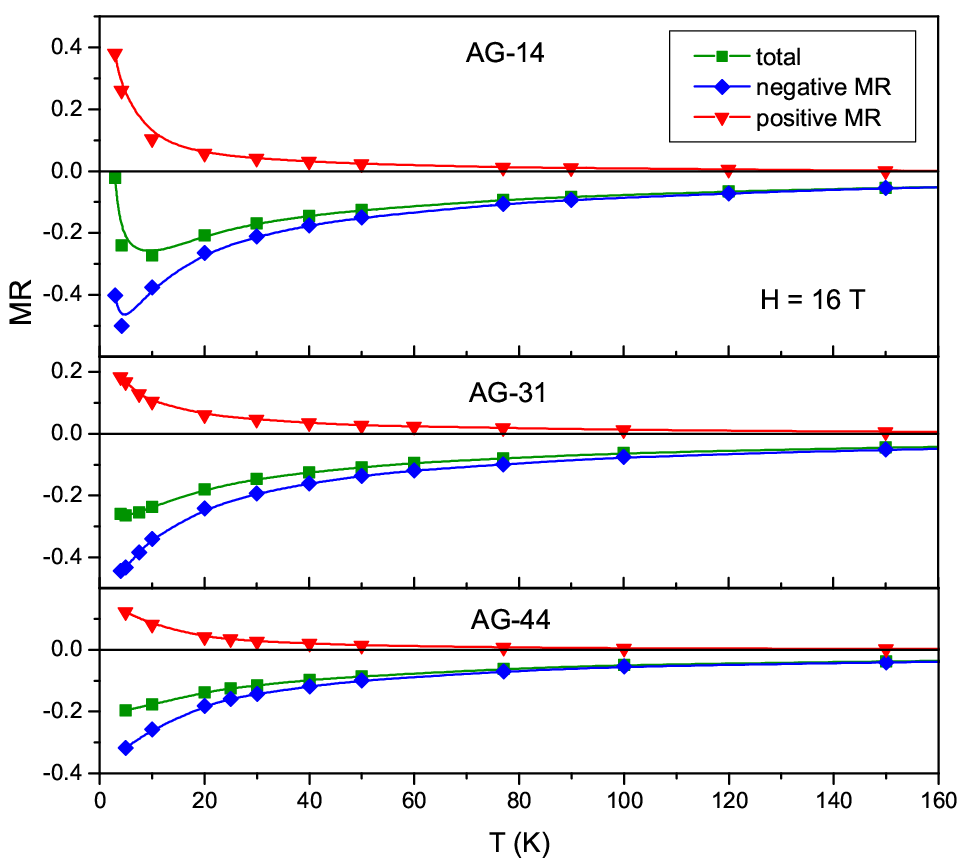}
  \caption{Temperature dependences of experimental values of the magnetoresistance and corresponding values of negative and positive contributions in magnetic field of 16~T for samples AG-14, AG-31 and AG-44.}
  \label{fig:MR_H16T_all}
\end{center}
\end{figure}

The ratio of the values of negative and positive contributions to the magnetoresistance at different temperatures in the maximum field of 16 T for the samples of three compositions is illustrated in Fig.~\ref{fig:MR_H16T_all}. It can be seen that as the carbon content (thickness of the graphene shell) increases, the magnitudes of both contributions decrease, and unlike the negative contribution, the positive contribution becomes insignificant at $T\gtrsim 100$~K with increasing temperature.

\subsubsection{Parameters of negative contribution to the magnetoresistance}

\begin{figure}[h]
\begin{center}
\includegraphics[width=8cm]{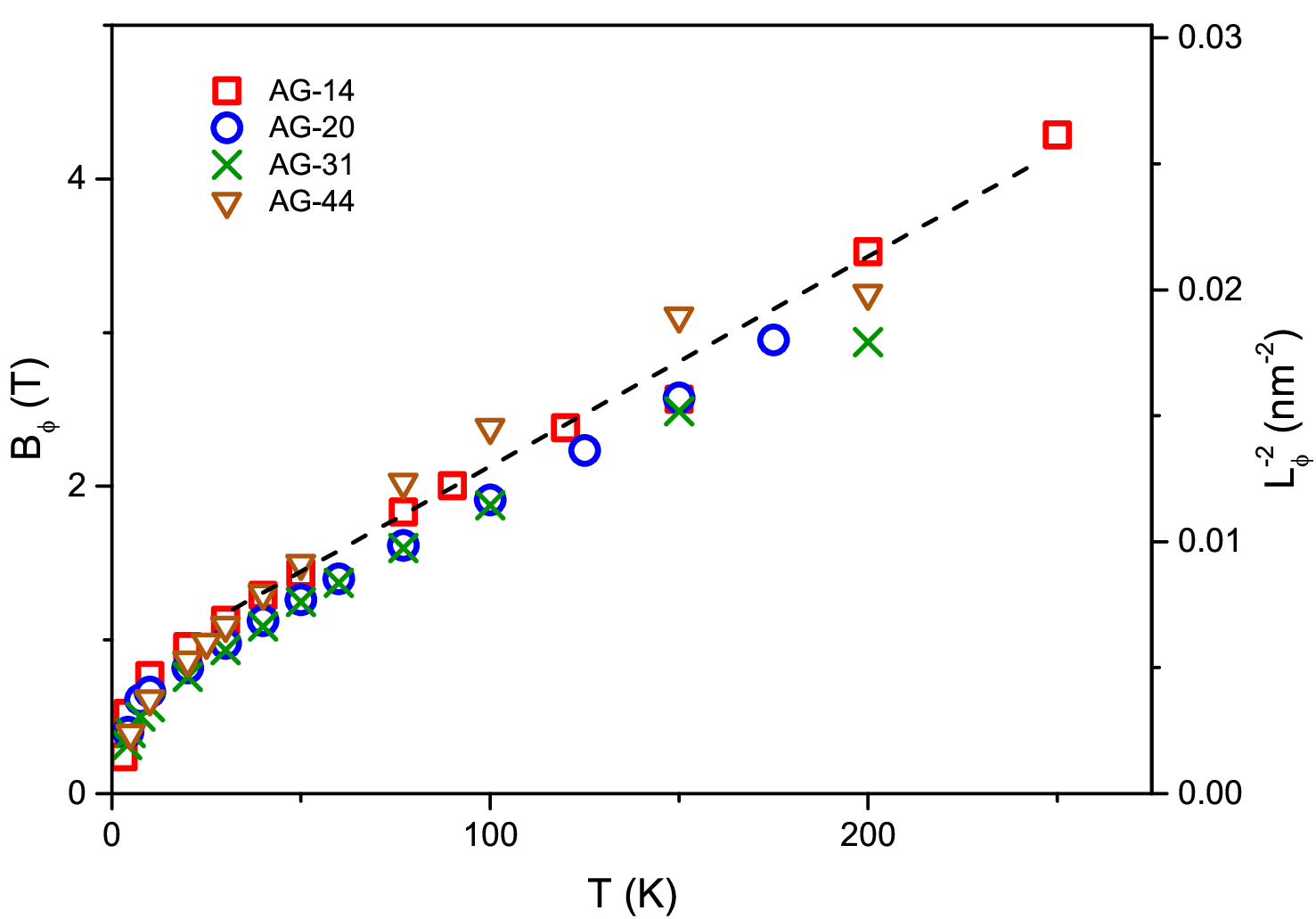}
  \caption{Temperature dependences of the parameter values $B_{\phi}$. The right axis shows the corresponding values $L_\phi^{-2}(T)$.}
  \label{fig:B_phi_all}
\end{center}
\end{figure}

The temperature dependences of the parameter $B_{\phi}$ in Eq.~(3) for samples of all compositions are grouped around a single dependence, which is linear at $T>20$~K (see Fig.~\ref{fig:B_phi_all}). Since the phase coherence length $L_{\phi}$ is related to $B_{\phi}$ by the relation $$L_{\phi}=\sqrt{\hbar/(4eB_{\phi})}$$ the value of $L_\phi^{-2}(T)$ also changes linearly in this temperature region (see Fig.~\ref{fig:B_phi_all}, right scale).

The phase coherence length is defined by the formula $$L_\phi=\sqrt{D\tau_\phi}\,,$$ where $D$ is the diffusion coefficient, $\tau_\phi$ is the dephasing time. As it is known, at weak localization the phase breaking of carriers motion at inelastic interactions can occur not only (and, in the case of graphene due to weak electron-phonon interaction, not so much) due to phonon scattering, but primarily due to electron-electron scattering \cite{altshuler82}. In this case, the expression for the temperature dependence of $\tau_\phi^{-1}$ in a 2D system can be written in the form \cite{aleiner99}
\begin{equation}\label{eq:tau-phi}
    \frac{1}{\tau_\phi}= \frac{k_B T}{\hbar}\frac{R_{\Box}e^2}{2\pi\hbar}
    \ln\left(\frac{\pi\hbar}{R_{\Box}e^2}\right),
\end{equation}
where $R_{\Box}$ is the resistance per square. In this case, $L_\phi^{-2}$ should depend linearly on temperature, which is observed in the samples of all four compositions, starting from 20~K and, practically, up to room temperature (see Fig.~\ref{fig:B_phi_all}).

\begin{figure}[h]
\begin{center}
\includegraphics[width=8cm]{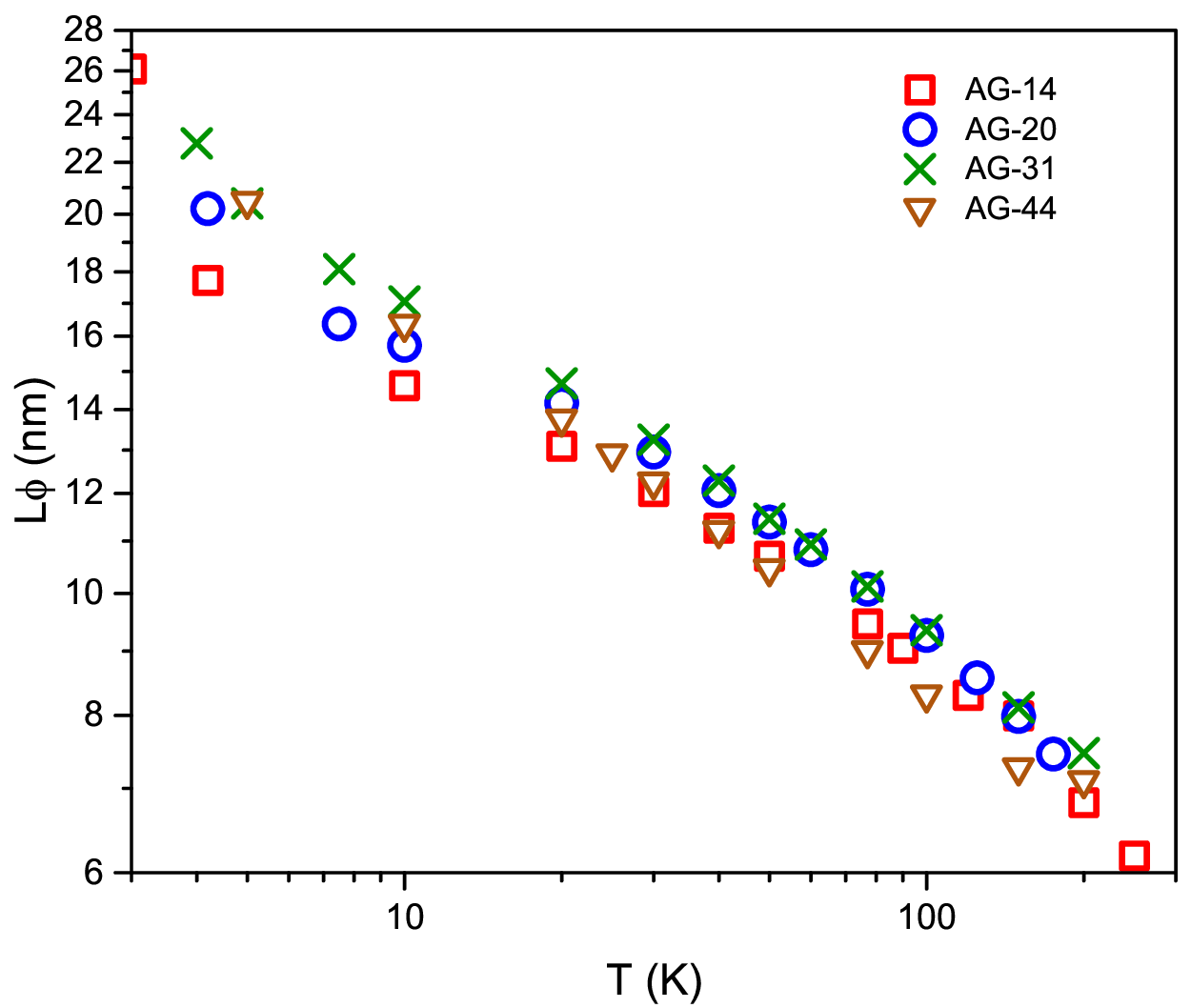}
  \caption{Temperature dependences of the phase coherence length $L_{\phi}(T)$, calculated from the values of parameter $B_{\phi}(T)$.}
  \label{fig:L_phi_all}
\end{center}
\end{figure}

Note that such a large negative magnetoresistance (up to 50\% at helium temperatures and about 2\% at room temperature) for all samples, together with a linear dependence $L_\phi^{-2}(T)$, extending to similarly high temperatures, is observed in a highly defective graphene system, apparently, for the first time. The main reason for these large magnetoresistance values is apparently the very high degree of defectivity of the graphene coating of the aerogel nanofibers. The coefficient $$A = 2\pi\hbar / R_{\Box}e^2$$ reaches the value of one for AG-14 at 4.2~K. Accordingly, this gives an estimate of $R_{\Box}\approx 8.1\cdot 10^4$~Ohm.

The extrapolation of the linear part of the $L_\phi^{-2}(T)$ dependence to zero temperature indicates that along with the temperature-dependent contribution, there is also a temperature-independent contribution to the phase coherence length. At the same time, the behavior of the $L_\phi^{-2}(T)$ (or, respectively, $\tau_{\phi}^{-1}(T)$) dependence at low temperatures on our samples is very different from that observed earlier on CVD- and epitaxial graphene samples with sizes of several tens of microns \cite{lara2011,baker2012} or on micron-sized graphene samples prepared by mechanical exfoliation \cite{tikhonenko2008,ki2008}. In all these studies, the values of $\tau_{\phi}^{-1}(T)$ at low temperatures reach a saturation state, approaching the value of the temperature-independent contribution $\tau_{\phi0}^{-1}(T)$. It was assumed in Refs.~\cite{lara2011,baker2012} that this contribution is related to ``spin-flop'' scattering on local magnetic moments, which may be possessed by some structural defects or impurities. It was shown in Ref.~\cite{tikhonenko2008} that the saturation yield occurs when the magnitude of $L_\phi$ is compared to the sample size. Also spatial inhomogeneity of carrier density can be a possible cause \cite{baker2012,ki2008}. In all these studies on weak localization in graphene, the values $L_{\phi0}$ were in the range of 0.6--2 $\mu$m. However, similar behavior of $L_\phi^{-2}(T)$ at $T\rightarrow 0$ was also observed in studies of magnetotransport in disordered multilayer carbon nanotubes \cite{piraux2015,tarkiainen2004} at much smaller values of $L_{\phi0}$ (the order of 20--40 nm).

In contrast to the above-mentioned works, in our case the dependence $L_\phi^{-2}(T)$ shown in Fig.~\ref{fig:B_phi_all} not only does not reach saturation with decreasing temperature, but, on the contrary, becomes stronger. A similar behavior was observed in Ref.~\cite{tikhonenko2009}, where a narrow strip of sufficiently perfect graphene obtained by mechanical exfoliation was studied. From the data presented in this work, it can be seen that below 40~K there is a more rapid decrease in the value of $\tau_{\phi}^{-1}$, which is proportional to $L_\phi^{-2}$. It should be noted that such character of the temperature dependence the authors observed only at a small number of carriers and in conditions of strong antilocalization. As can be seen from Fig.~\ref{fig:B_phi_all} of our work, this behavior of $L_\phi$ takes place at temperatures when the value of $L_\phi$ becomes of the order of the average aerogel filament diameter (15~nm) \cite{tsebro2022}. It is reasonable to assume that when the temperature drops below 15--20 K, a gradual change in the conduction regime from two-dimensional to one-dimensional begins to occur (transition 2D $\rightarrow $ 1D).

\begin{figure}[h]
\begin{center}
\includegraphics[width=8cm]{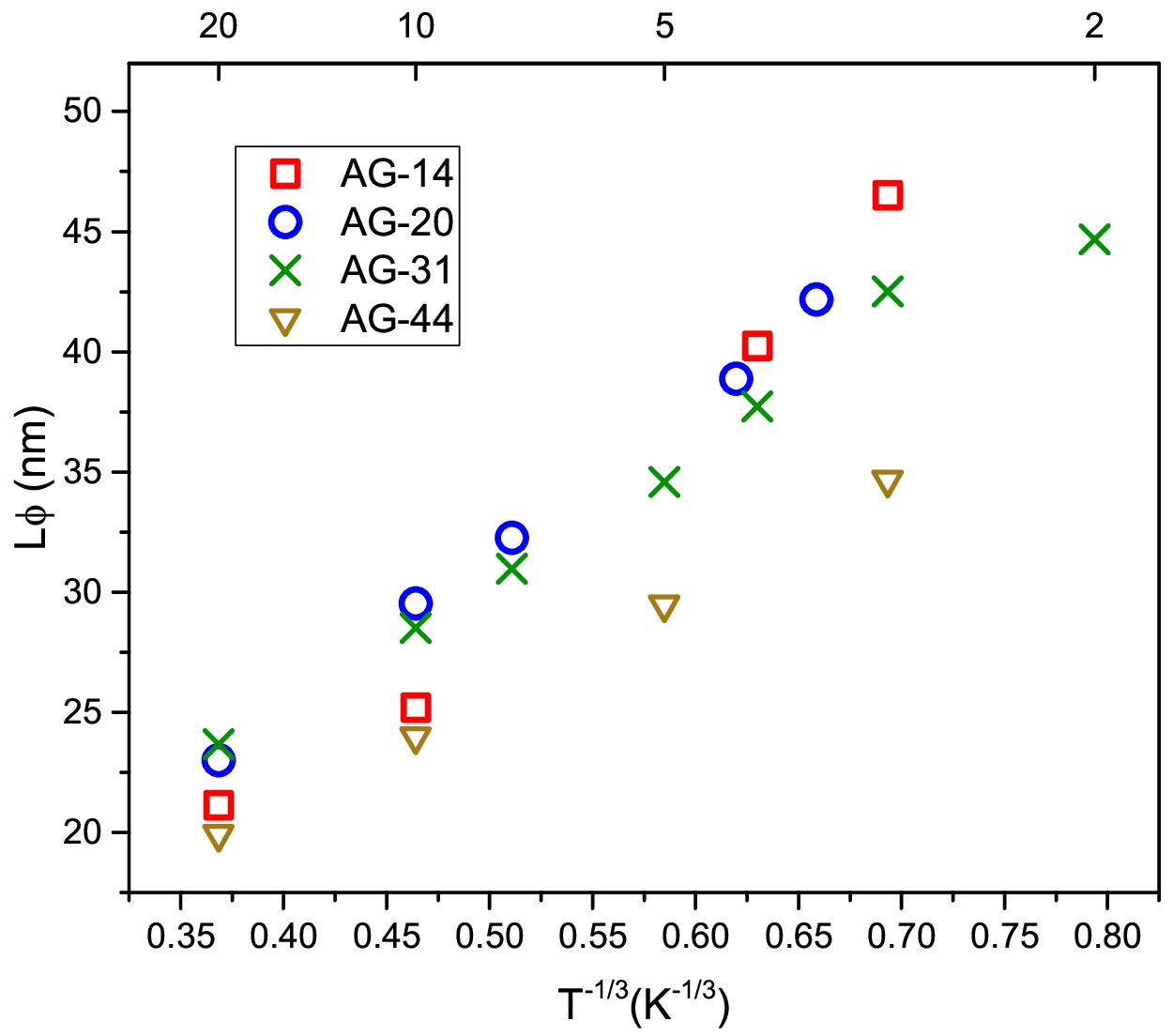}
  \caption{Dependences of $L_\phi(T)$ on $T^{-1/3}$ based on the results of approximation of field dependences of magnetocondactance in one-dimensional model.}
  \label{fig:L_phi_1D}
\end{center}
\end{figure}

In the one-dimensional case, the expression for the temperature dependence of $\tau_\phi^{-1}$ has the form \cite{aleiner99}
\begin{equation}\label{eq:tau_phi_1d}
    \frac{1}{\tau_\phi}= \left(\frac{k_B T e^2\sqrt{\hbar\,D}}{\hbar\, \sigma_1}\right)^{2/3},
\end{equation}
where $\sigma_1$ is the conductivity of a unit length of a one-dimensional conductor, and the quantum correction to the conductivity is described by the following expression
\begin{equation}\label{eq:corr_1d}
    \delta\sigma_{1\rm{D}}=-\frac{e^2L_{\phi}}{\pi\hbar}\frac{1}
    {[-\ln \rm{Ai}(\tau_{\phi}/\tau_B)]'}\ ,
\end{equation}
where Ai is the Airy function, $\tau_B=3\hbar^2/D(eBa)^2$, $a$ is the transverse dimension of the one-dimensional conductor.

It turned out that the dependences of magnetodactance on the magnetic field measured by us at low temperatures (2--20 K) admit approximation by the
formulas of this one-dimensional model. The values $L_\phi(T)$ obtained from this approximation are proportional to $T^{-1/3}$ in accordance with expression (\ref{eq:tau_phi_1d}) (see Fig.~\ref{fig:L_phi_1D}) and exceed the corresponding results of the 2D-WL model (at 10 K about two times, see Figs.~\ref{fig:L_phi_all} and \ref{fig:L_phi_1D}). This can be considered a quite satisfactory agreement, taking into account that in the transition region neither model, strictly speaking, performs well. Qualitatively, the transition to a faster growth of $L_\phi$ with decreasing temperature at the 2D $\rightarrow $ 1D transition can be explained by the fact that there is a decrease in the number of possible states into which the electron can scatter, not only due to a decrease in the width of the interval $k_BT$, but also due to the fact that states near momentum directions perpendicular to the filament axis are eliminated from the ``game''. As for the temperature-independent contribution $L_{\phi0}^{-2}$, as can be seen from Fig.~\ref{fig:B_phi_all}, it decreases significantly (and $L_{\phi0}^{-2}$, respectively, grows) if we extrapolate to points below 20~K. Whether this is related to the 2D $\rightarrow $ 1D transition remains questionable.

\subsubsection{Positive contribution to magnetoresistance}

\begin{figure}[h]
\begin{center}
\includegraphics[width=8cm]{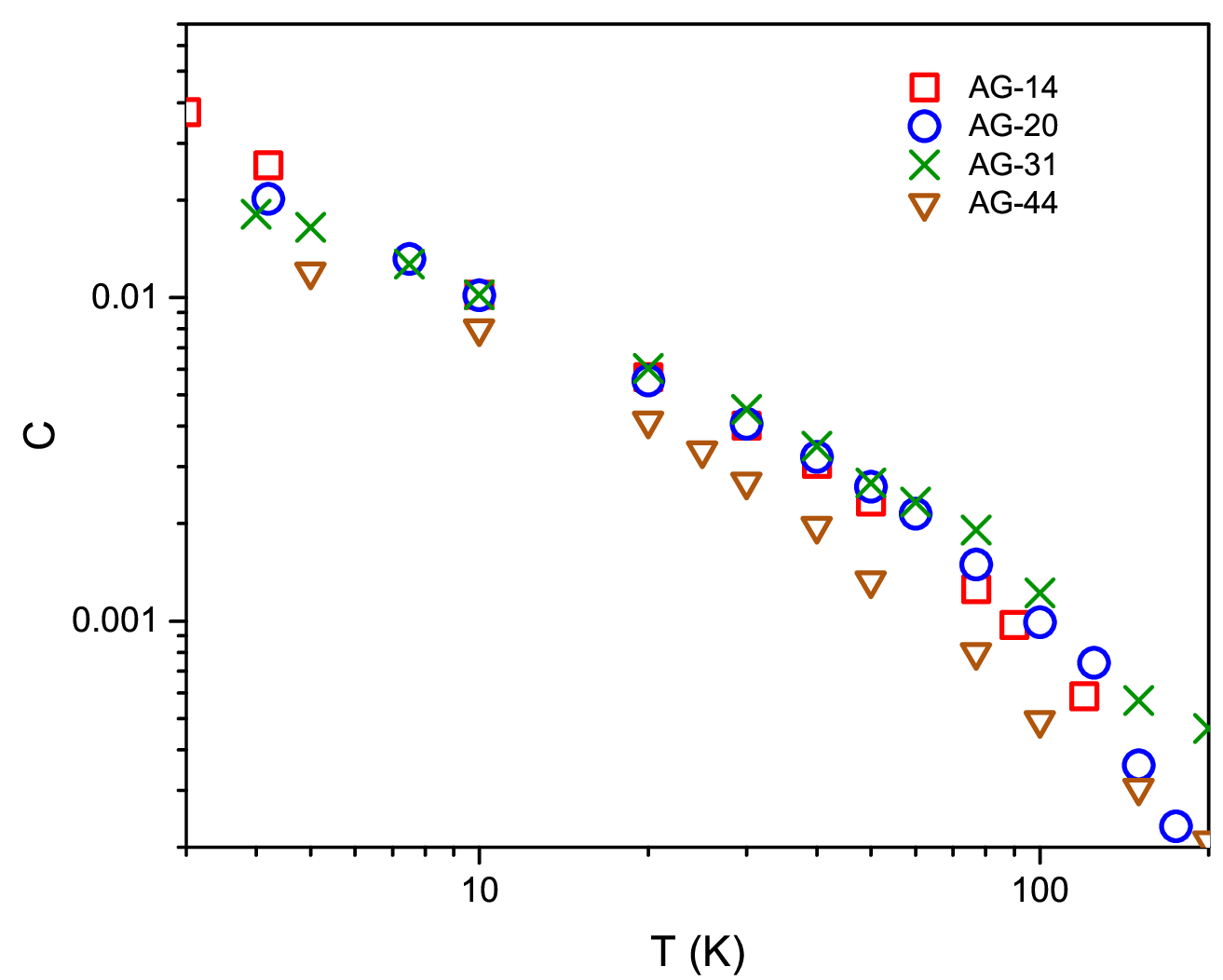}
  \caption{Temperature dependences of the coefficient $C$ in Eq.~(\ref{eq:MR_plus}) of the positive linear contribution to the magnetoresistance.}
  \label{fig:C1_all}
\end{center}
\end{figure}

Fig.~\ref{fig:C1_all} shows the temperature dependence of the coefficient $C$ in Eq. (5), which characterizes a linear positive contribution to the magnetoresistance, unsaturated in fields up to 16 T. This contribution manifests itself primarily at low temperatures (see Fig.~\ref{fig:AG-14_T4K_3in1}) and becomes insignificant at $T\gtrsim 100$~K (see Fig.~\ref{fig:MR_H16T_all}).

Linear nonsaturating magnetoresistance (LNMR) in systems with strong inhomogeneity has recently been the subject of many publications (see, for example, Refs.~\cite{rama2017,ping2014,naray2015,zhu2022,friedman2010,kisslinger2015,gu2021}). Attention to the LNMR effect was drawn after experimental studies \cite{xu97,husmann2002} on LNMR on silver chalcogenides, as well as theoretical studies \cite{parish2003,parish2005}. In the latter, it was shown by numerical modeling on a two-dimensional random resistor network (RRN) that taking into account the Hall contribution to the potential distribution on the grid leads to the LNMR effect in a magnetic field perpendicular to the grid plane. This result was proposed as an explanation of the results of \cite{xu97,husmann2002}, and this explanation was accepted by the authors of many subsequent experimental studies, where the LNMR effect was observed for a wide variety of objects, including single-layer CVD graphene \cite{ping2014}, bilayer mosaic graphene \cite{kisslinger2015}, multilayer epitaxial graphene \cite{friedman2010}, and such objects as, for example, layered compounds of transition metal dichalcogenides \cite{zhu2022,gu2021}.

Another approach to the explanation of LNMR (the so-called effective medium theory) was proposed in Ref.~\cite{guttal2005}, where it is shown that this effect in an inhomogeneous conductor should be observed when it consists of different volume regions of electrons and holes with homogeneous conductivity. The equivalence of the effective medium and RRN models is demonstrated in \cite{rama2017} using data from several experimental works for different objects. Both these models lead to the same results and along with the linear dependence of magnetoresistance in strong fields predict a quadratic dependence at $H\rightarrow 0$.

In our case, taking into account the fact that the conducting medium of graphenized samples of nematic aerogel, namely, the bound, strongly stretched along one direction, chaotic in cell size mesh of graphene shells of mullite filaments, is by no means a homogeneous medium, the above explanation of the linear contribution to magnetoresistance seems to be quite justified. It should be noted that this contribution, unlike other magnetoresistance mechanisms, is not related to any particular type of conductivity, such as diffusion or hopping. At the same time, it can be noted that the characteristic size of inhomogeneities associated with this contribution is much larger than the characteristic sizes essential for this or that type of conductivity (free path length, localization length), i.e., it has a mesoscopic order of magnitude.

\subsection{Temperature dependence of resistance. Hopping conductivity}

\begin{figure}[h]
\begin{center}
\includegraphics[width=8cm]{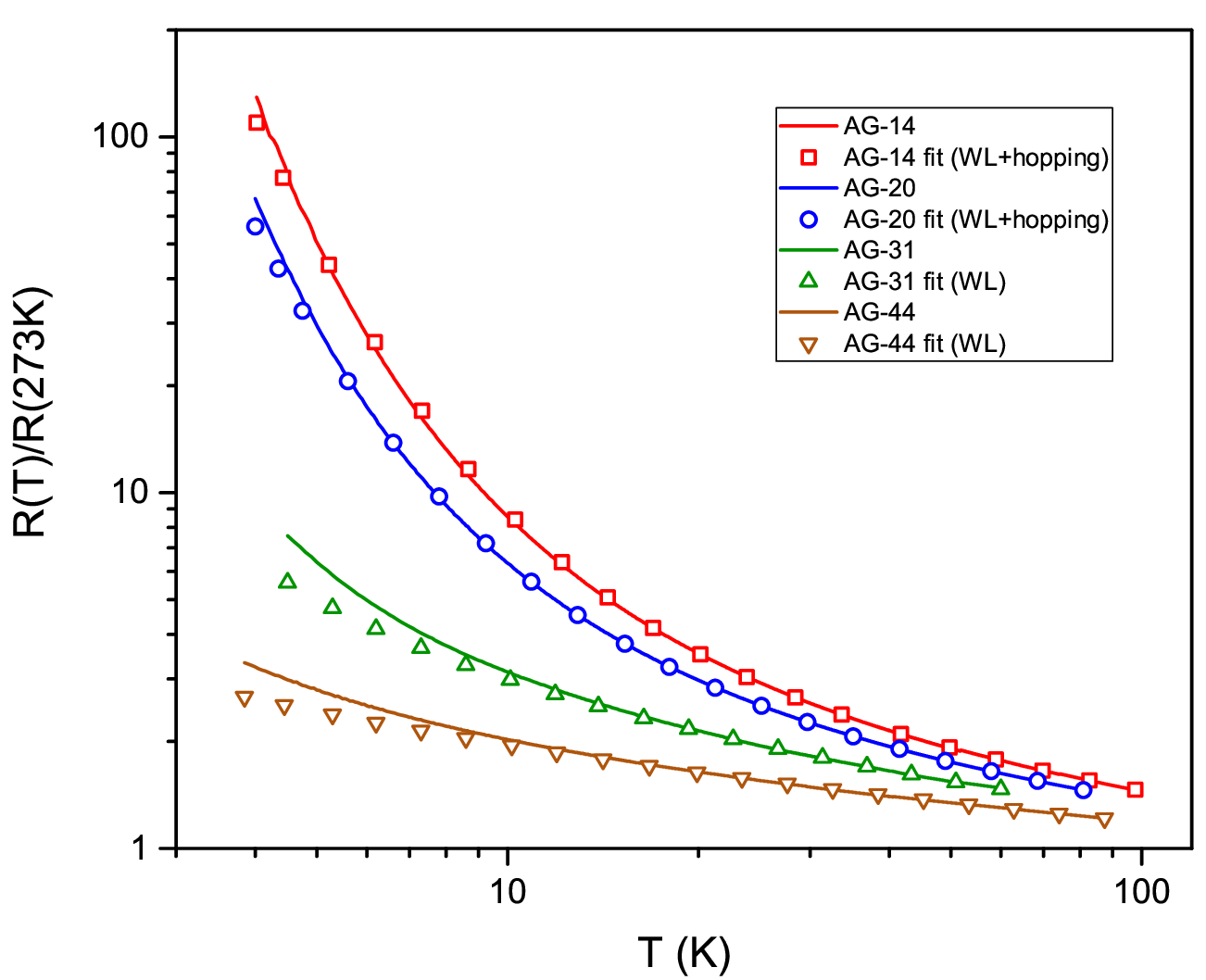}
  \caption{Normalized temperature dependences of the resistivity of aerogel samples of all compositions. Symbols show the results of approximation according to Eq.~(\ref{eq:approx_R}).}
  \label{fig:R(T)_all}
\end{center}
\end{figure}

Fig.~\ref{fig:R(T)_all} shows the temperature dependences of resistivity measured in a zero magnetic field for aerogel samples of all compositions in the temperature region ($T < 100$~K), where the resistivity varies most significantly. The dependences of the resistance normalized to the value at $T=273$~K are shown, which excludes the geometric factor and allows us to compare the relative values of resistivity as a function of composition and temperature. From the given data we can see that the range of resistance variation is significantly different for samples with minimal thickness of graphene shell (AG-14, AG-20) and samples with greater thickness (AG-31, AG-44).

Assuming that (a) mainly the conductivity of highly defective graphene nanofiber shells is determined by diffusive transport across graphene grains under weak localization conditions when the conductivity $\sigma(T)\propto \ln T$, and (b) at small thicknesses the continuity of the graphene coating may be broken and hopping conductivity between separate fragments of the graphene shell may occur, for samples AG-14 and AG-20 we approximated the experimental dependences $R(T)$ by the sum of two contributions with four fitting parameters $a,b,c$ and $T_0$.
\begin{equation}\label{eq:approx_R}
    R(T)/R(273{\textrm{K}}) = R_{\textrm{WL}}(T) + R_{\textrm{hop}}(T)\ ,
\end{equation}
where
\begin{equation*}
    R_{\textrm{WL}}(T) = 1/(a + b\ln T)\,,
\end{equation*}
\begin{equation*}
    R_{\textrm{hop}}(T) = c\ \exp \left[ \left(\frac{T_0}{T}\right)^{1/2} \right]\,.
\end{equation*}

The results of such approximation are shown by symbols in Fig.~\ref{fig:R(T)_all}.

The choice of $\alpha = 1/2$ for the hopping contribution in Eq.~(\ref{eq:approx_R}) in the procedure of approximation of experimental dependences $R(T)$ for samples AG-14 and AG-20 is due to the fact that this law is characteristic for granular conductors \cite{zhang2004,beloborod2007}, as well as for the mechanism of hopping conductivity between nanowires in the insulator matrix, considered in \cite{hu2006a,hu2006b}. The values of the parameter $T_0$ obtained for samples AG-14 and AG-20 as a result of approximation $R(T)$ are equal 260 and 210~K, respectively.

For samples AG-31 and AG-44 with a larger thickness of the graphene nanofiber shell, the graphene coating appears to be continuous. For these samples, the relative change in resistance at $T < 100$~K is small, and the single contribution of $R_{\textrm{WL}}$ in Eq.~(\ref{eq:approx_R}) appeared to be sufficient to describe it. Taking into account the fact that below 15--20~K begins the transition 2D $\rightarrow $ 1D, approximation of experimental dependences $R(T)$ by the first term of the sum~(\ref{eq:approx_R}) was carried out from 15~K and above, and below this temperature, as can be seen from Fig.~\ref{fig:R(T)_all}, the deviation of experimental curves upward from the approximation values correlates with the previously stated assumption about the change at lowering the temperature of the conduction regime from two-dimensional to one-dimensional. For samples AG-14 and AG-20 such deviation is not visible against the background of strong hopping contribution.

Thus, the hopping transport in the studied graphene aerogels with the value of $\alpha = 1/2$ in Eq.~(\ref{eq:mott}) explicitly manifests itself only for samples with the minimum thickness of the graphene shell of the aerogel nanofibers.

\section{Conclusion}

Thus, in the present work, the magnetotransport properties of nematic aerogels consisting of graphene-coated nanofibers Al$_2$O$_3\cdot$SiO$_2$ have been investigated in the temperature range from 3 to 300~K and magnetic fields up to 16~T. It is shown that the measured magnetoresistance is well enough approximated by the sum of two contributions --- negative, described in the framework of the 2D model of weak localization, and linear in the field positive, unsaturated in strong magnetic fields.

It was found that for all the studied samples the value $L_\phi^{-2}$ obtained from the analysis of negative magnetoresistance above 20~K depends linearly on temperature, which indicates electron-electron scattering as the main mechanism of dephasing. At lower temperatures, $L_\phi^{-2}$ decreases faster than linearly. Since in this temperature
region the phase coherence length $L_\phi$ becomes comparable to the diameter of the nanofibers, it can be assumed that this behavior is related to the onset of the transition below 20 K from a two-dimensional localization regime to a one-dimensional regime, where $L_\phi^{-2}$ is proportional to $T^{2/3}$. The approximation of the magnetoresistance below 20~K using the formula for weak 1D localization gives values $L_\phi$ that only exceed twice the corresponding values obtained within the 2D model.

We note the large magnitude of the negative magnetoresistance (up to 50\% at helium temperatures and about 2\% at room temperature) together with the linear dependence $L_\phi^{-2}(T)$, extending to the highest temperatures. This appears to be observed for the first time in a highly defective graphene system. The reason for this is apparently the very high degree of defectivity of the graphene coating of the aerogel nanofibers.

The linear positive contribution to the magnetoresistance can be explained by the inhomogeneous distribution of the local charge carrier density in the system, which leads to the admixture of the Hall component to the longitudinal component of the bulk resistivity tensor.

The temperature dependence of the resistivity of nematic aerogels in the zero field can also be represented as a sum of two contributions, one of which corresponds to regions with diffusion character of transport and is described by the expression for the case of weak 2D-localization. For the second contribution, the formula for variable range hopping conductivity with the exponent equal to 1/2 (Efros-Shklovsky law) was used. For samples with high carbon content the second contribution is negligibly small and below 10~K a significant deviation from the theoretical dependence is observed, which also confirms the assumption about the transition below this temperature to the one-dimensional regime of weak localization.

\section*{Acknowledgments}

The authors are grateful to B.I. Shklovsky for useful discussions of the peculiarities of hopping transport in the studied system.

\end{document}